\documentclass[12pt]{article}
\usepackage[english]{babel}
\usepackage[utf8]{inputenc}
\usepackage{amsmath}
\usepackage{natbib}
\usepackage{graphicx}
\usepackage{caption}
\usepackage{subcaption}
\usepackage[margin=1in]{geometry}
\usepackage{array}
\newcolumntype{L}[1]{>{\raggedright\let\newline\\\arraybackslash\hspace{0pt}}m{#1}}
\newcolumntype{C}[1]{>{\centering\let\newline\\\arraybackslash\hspace{0pt}}m{#1}}
\newcolumntype{R}[1]{>{\raggedleft\let\newline\\\arraybackslash\hspace{0pt}}m{#1}}
\date{}

\begin{document}
\title{An Overview of Drone Energy Consumption Factors and Models}

\author{\small Pedram Beigi$^1$ \and \small Mohammad Sadra Rajabi$^2$ \and \small Sina Aghakhani$^3$}
\date{\footnotesize%
    $^1$Department of Civil Engineering, Sharif University of Technology, Tehran, Iran\\%
    $^2$School of Civil Engineering, University of Tehran, Tehran, Iran\\%
    $^3$Department of Industrial Engineering, Sharif University of Technology, Tehran, Iran\\[2ex]%
}

\maketitle

\begin{abstract}
At present, there is a growing demand for drones with diverse capabilities that can be used in both civilian and military applications, and this topic is receiving increasing attention. When it comes to drone operations, the amount of energy they consume is a determining factor in their ability to achieve their full potential. According to this, it appears that it is necessary to identify the factors affecting the energy consumption of the unmanned air vehicle (UAV) during the mission process, as well as examine the general factors that influence the consumption of energy. This chapter aims to provide an overview of the current state of research in the area of UAV energy consumption and provide general categorizations of factors affecting UAV's energy consumption as well as an investigation of different energy models.
\end{abstract}

\hfill\break%

\maketitle
\section{Introduction}

Drones offer a number of advantages over trucks while they are more efficient. They eliminate the need of the drivers and can often travel with a higher speed than vehicles since they are not restricted to the road systems \citep{agatz2018optimization}. These advantages enable logistics companies and online retailers to deploy drones to deliver packages quickly. Humanitarian organizations are also actively considering using drones in disaster situations \citep{cheng2020drone}. Also, drones have a significant environmental advantage over trucks by reducing emissions. While UAVs have a number of desirable features, the limited battery life is a major limitation for them. As most UAVs are electric devices powered by onboard batteries, this constraint significantly limits their capabilities. Many studies have been recently proposed contributing towards saving energy and increasing endurance. A major focus of these contributions is the design of an automated system for charging and replacing batteries \citep{yacef2017optimization}. The smaller UAVs do not entirely solve the mechanization problem since they have one major flaw, which is insufficient power \citep{alwateer2019enabling}. Larger drones, such as those primarily employed in military applications, have enough power sources, but this advantage comes at the cost of being considerably larger, less maneuverable, and rather loud. The importance of having an appropriate power source is critical since it leads to lengthy flight endurance. It would allow further flight mission planning and optimal recharging for UAVs. It is therefore essential to plan and design UAV missions in an energy-efficient manner. In order to achieve this, a reliable power consumption model is required for predicting the power consumption \citep{abeywickrama2018empirical}.

\section{Factors Affecting Energy Consumption of Drones}

The features and configurations of UAVs vary considerably depending on their missions. Understanding the elements of the determined energy use is critical for designing energy consumption models that are accurate and efficient. Drone activities are more energy-sensitive than conventional vehicle operations \citep{cheng2020drone}. Internal and external factors can affect energy consumption. As an example, the lower power consumption was observed when flying into headwinds \citep{tennekes2009simple}, which is due to the increasing thrust generated by the translational lift as the UAV moved from hovering to forward flight. Temperature and air density are also linked to battery drain and lift capacity of aircraft. Below zero degrees Fahrenheit, UAVs fly shorter distances and experience more malfunctions. The weight and payload of UAVs also individually affect their energy consumption more than all other factors \citep{thibbotuwawa2018energy}. There is an analysis of different parameters that influence the energy consumption of the UAV Routing Problem in \citep{thibbotuwawa2018factors}. This is achieved by analyzing an example scenario of a single UAV multiple delivery mission. Based on the analysis, the relationships between UAV energy consumption and the influencing parameters are examined. Therefore, it is vital to have better knowledge and estimate of drone energy use. The four main elements that influence drone energy usage are drone design, environment, drone dynamics, and delivery operations, which will be discussed further. \citep{demir2014review, zhang2021energy}.

\subsection{Drone Design}

The weight and size of the drone's body, number and size of rotors, weight, size, and energy capacity of the battery, power transfer efficiency, maximum speed and payload, lift-to-drag ratio, delivery mechanism, and avionics are all elements to consider while designing a drone \citep{zhang2021energy}. It is inherently complex to design mechatronics systems since they involve multiple domains. During the design process, the different engineering domains involved in the activity influence one another, which makes the task of designing a complex process for design engineers \citep{mohebbi2014trends}. The mechatronics systems are traditionally designed sequentially, with the mechanical component coming first, followed by the electronic components, then the control strategy. 

In order to achieve an optimized design, the coupling between the different components and domains must be evaluated in the early stage of the design process to avoid negative consequences associated with dependency \citep{Alyaqout2011combined}. Several methods have been suggested in order to achieve a better design that incorporates both mechanical and control aspects of the mechatronics system. The proposed methods tend to aim an optimal aspect of the system, for example, the control or the mechanisms in isolation. The literature has identified some approaches for the design support of drones or mechatronics systems in general. As an example, the design for control strategy applied to visually served drones is presented by \cite{mohebbi2015integrated}. This process involves simplifying the dynamics model of the system in order to better understanding and improving its representation and then devising a control algorithm that will enhance the control of system. A further method for designing a structure-control system is a robust structure-control design, which uses nonlinear dynamic multi-objective optimization to design a system. It considers the interactions between the structure and the control to propose a robust design, as presented by \cite{Alyaqout2011combined}. This method relies on the design of the controller to achieve the robustness of the system, which limits it to a robust approach of the control. For both of these methods, it is the control component on the focus, and little information is gleaned about the mechanisms; in addition, the interaction goes in a single direction, from control to mechanism, while the other direction can only be achieved by further simplification of the dynamics by adding extra constraints, such as stability criteria \citep{mohebbi2015integrated}. \cite{coulombe2017applying} aims to develop a robust design for a quadrotor drone, with particular attention to structural parameters, such as the mass and dimensions. In definition, a robust design method is one that emphasizes the minimization of the effects of variation in design parameters on the response of the system. This paper presents a system's response in terms of its energy consumption. Monte Carlo simulation is used to determine the most influential design parameters, and then a designer-defined objective function is minimized to determine a robust mechanical design for the quadrotor under consideration.

\subsection{Environment}

Air density, gravitational force, wind conditions, weather (snow, rain, etc.), ambient temperature, and operational restrictions are all environmental elements \citep{zhang2021energy}. The existing research indicates that reduced power consumption was observed when flying into headwinds \citep{tennekes2009simple}, which can be attributed to the increasing thrust caused by translational lift. In the presence of a headwind, the translational lift will increase as the relative airflow increases, resulting in reduced power consumption for hovering. If the wind speed exceeds a certain threshold, aerodynamic drag may outweigh the benefits of translational lift \citep{alyassi2022autonomous}. In addition, temperature and air density have a relationship, which is linked to the battery drain. The air density around a flying aircraft changes with temperature, thus affects their lift capacity. Studies have been shown that UAVs tend to fly shorter distances and experience increased malfunction rates in cold weather conditions (below zero degrees Fahrenheit). Outdoor routing for UAVs must account for the stochasticity of weather variables that affect UAV energy consumption \citep{kinney2005devising}. The majority of UAV routing studies either assume infinite fuel capacity or presume that they would never run out of fuel \citep{frazzoli2004decentralized} or do not take fuel into consideration \citep{thibbotuwawa2018energy}. The weather's impact on UAV routing is influenced by two primary elements, which are: i) Wind: the major environmental factor that affects the UAV is wind direction and its speed. In some circumstances, wind can reduce energy consumption while increasing resistance to movement.
ii) Temperature: since temperature is connected to battery drain and capacity, it might impair the UAV's battery performance \citep{dorling2016vehicle}.

Ignoring the effects of the weather will not result in more realistic answers \citep{erfani2020risk}, as flying with the wind can cut energy usage, and cold temperatures can harm battery performance \citep{dorling2016vehicle}. 
As weather changes over time in a stochastic manner \citep{wu2014autonomous}, one must expect that the fuel consumption of a particular route will vary at different times \citep{thibbotuwawa2018energy}. It is critical that the drone be highly mobile and unaffected by the surroundings \citep{tang2015drone}.

\subsection{Drone Dynamics}

Drone dynamics factors include drone travel speed, drone motion (i.e., takeoff/landing, hover, horizontal flight), acceleration/deceleration, angle of attack, and flight altitude. The idea of drones being transported for part of their journey on other vehicles also should be considered, such as trucks or public transportation \citep{zhang2021energy}.  Different cargo weights can have a major influence on energy consumption models; thus they should be taken into account. \citep{alyassi2017autonomous, dorling2016vehicle}. Fuel/energy usage is recognized to be dependent on various aspects in the airline industry. Take-off gross weight, empty weight, and thrust to weight ratio might limit the maximum flying distance or time of a UAV \citep{shetty2008priority}, fuel weight, and payload \citep{zhang2015space}. 
Wind speed and direction are related to flying speed since the direction of the wind can affect the UAV's flying status either positively or adversely. A UAV's flight status might be one i) hovering; ii) horizontal moving or cruising or level flight; iii) vertical moving: vertical take-off/landing/altitude change.
As a result, while calculating energy consumption, the UAV's flying state, as well as its speed, should be taken into account \citep{alyassi2017autonomous}. During a drone delivery journey, many of these variables are interrelated and dynamic. 

All of these elements, notably drone design, drone dynamics, and delivery procedures, might cause uncertainty in calculating drone energy usage. 

\subsection{Delivery Operations}

Delivery operations factors may only apply to drone delivery missions, whereas other missions require consideration of other factors, which is not mentioned here.  Weight and size of the payload, "empty returns" (i.e., the return trip after delivery is without the payload, implying a successful delivery), fleet size and mix, number of deliveries per trip, delivery mode, and service region area are all crucial factors in a delivery operation. Some of these elements are specified or limited by the drone design (e.g., maximum payload, the projected area of the drone, etc.), while others are operational parameters \citep{moeinifard2022lost, aghakhani2022new} that can vary for a given drone design (e.g., payload, speed, etc), and still others are external factors (e.g., weather) \citep{zhang2021energy}. As drone delivery continues to develop, many literature have been considered the use of drones in transportation problems. \cite{dorling2016vehicle} investigated vehicle routing problems for drone delivery (VRPDD). Based on an energy consumption model for drones, the authors investigated the implications of payload and battery weight on energy consumption. \cite{dukkanci2021minimizing} describes the Energy Minimizing and Range Constrained Drone Delivery Problem (ERDDP), where drones are used to deliver products to a number of customers, and the drones themselves are transported by traditional vehicles. As part of the ERDDP, (i) launch points will be selected among a possible set of sites from where drones will take off to serve a number of customers, (ii) customers will be assigned to the launch points, and (iii) the speed at which drones will travel between the launch points and the customers will be determined. It proposes a nonlinear model for ERDDP that minimizes the total operational cost, including an explicit calculation of the drone's energy consumption in relation to its speed. The results demonstrated the effect of various factors on location, assignment, and speed decisions. As a result of two problems related to drone-assisted cargo delivery (flying sidekick traveling salesman problem (FSTSP), parallel drone scheduling to travel salesmen (PDSTSP)) based on \cite{murray2015flying}, the researchers concluded that the speed of a drone has a significant effect on drone delivery operations due to range alternation. Even if the drone range is reduced, it is preferable to have higher speeds. Other studies used the FSTSP setup have examined coordination between trucks and drones (e.g., \cite{wang2017vehicle, poikonen2017vehicle, carlsson2018coordinated}) with some extending the problem to minimize operational costs.

\section{Modeling of Drone Energy Consumption}

Due to the limited energy provided by the lithium polymer (LiPo) batteries, which are typically installed on mini drones, the energy consumption of each drone plays a critical role in determining its figure of merit \citep{famili2022optilod}. To account for the drone's limited battery capacity, numerous algorithms have been proposed in the literature to assist in optimizing the different aspects of the energy considerations. Nevertheless, the most studies do not analyze the battery according to its actual performance \citep{chen2018case}. The most drone travel models impose time and/or distance limits. Some of the studies assume that the energy consumption is constant per unit of time or traveled distance; hence the drone energy consumption is modeled as a linear function of time or traveled distance (e.g., \cite{ferrandez2016optimization, ha2018min, huang2020reliable, moore2019innovative}). Also many models assumes the motor draining power at a 1:1 ratio to the battery. However, this is not correct since a battery supplies power with different efficiency values depending upon its state of charge (SOC).

Many models have been proposed in the literature with respect to the energy consumption; they consider a variety of parameters, aspects and missions such as optimal path planning, path following control, battery-aware and battery performance, target tracking, UAV-enabled mobile edge computing, UAV-enabled multicasting and drone's component models. The following subsections provide a review of these models.

\subsection{Optimal Path Planning}

In order to use UAVs in an optimal manner, path planning is one of the most important factors that can be realized in autonomous control. Path planning is a challenging process due to the increased number of parameters, such as control points, radar coverage areas, physical obstacles, etc., \citep{sonmez2015optimal}.
There are many methods that have been used to solve NP-hard optimal path planning problems, including heuristic methods (e.g., Christofides, Concorde) and meta-heuristic methods (e.g., genetic algorithms or discrete particle swarm optimization) \citep{wai2019adaptive}.

\cite{cheng2020drone} investigated a multi-trip drone routing problem (MTDRP) with time windows, where drones' energy consumption is modeled as a nonlinear function of payload and travel distance. Instead of using a linear approximation, they added logical cuts and subgradient cuts to the solution process in order to handle the more complex nonlinear (convex) energy function. A Branch-and-Cut algorithm is developed using a 2-index formulation.
\cite{wai2019adaptive} examined an optimal path planning and disturbance rejection control for a UAV surveillance system. A K-agglomerative clustering method is used to create a clustered 3D real pilot flight pattern and is then processed using A-star and set-based particle swarm optimization (S-PSO) algorithms to generate an optimal path planning scheme. The online adaptive neural network (ANN) controller combines a variety of learning rates with a fast disturbance rejection response to ensure control stability.
The traveling salesman problem (TSP) is used in \citep{zeng2018trajectory} in order to design the trajectory of an UAV so as to minimize the time taken to complete the mission in UAV-enabled multi-casting systems. \cite{sambo2019energy} applied a genetic algorithm to design a trajectory that consumes the least energy to visit all base stations and return to the UAV station.
\cite{van2021real} proposed optimal path planning approaches for UAVs to minimize their completion time and total energy consumption during data collection. A real-time optimization algorithm provides low computational complexity with fast deployment and low processing time for tracking and collecting sensor data. \cite{deng2022vehicle} aimed to develop a new vehicle-assisted UAV delivery solution that the energy consumption takes into account. It allows UAVs to serve multiple customers on a single take-off. In order to allocate tasks among UAVs and to plan the path of a vehicle, a multi-UAV task allocation model and a vehicle path planning model were developed. The model also considered the impact of changing the payload of the UAV on energy consumption, bringing the results closer to reality. To solve the problem, a hybrid heuristic algorithm based on an improved K-means algorithm and ant colony optimization (ACO) was proposed. \cite{thibbotuwawa2019planning} developed an off-line solution to the problem of UAV mission planning, taking energy consumption constraints and collision avoidance into account based on historical data. A predictive strategy is employed to avoid collisions between UAVs over the time horizon to create collision-free routing and schedules. \cite{morbidi2016minimum} discussed the problem of the minimum energy path through a model for the brushless DC motors and solved it with regard to the angular acceleration of the propellers of a quadrotor. \cite{dorling2016vehicle} analyzed the routing optimization for drone delivery services; however, the power model included only the weight of the battery in addition to the payload. The authors of \citep{di2015energy} presented an algorithm that minimizes the total energy consumed by the IRIS quadrotor, through a power model that describes the drone's energy consumption in different operating conditions. In \citep{abdilla2015power}, the authors studied the performance of different LiPo batteries and the models considered for battery runtime are based on the capacity rate effect, as well as Peukert's law \citep{di2015energy}.

\subsection{Path Following Control}

The trajectory control problem, defined as making a vehicle follow a pre-established path in space, can be solved by means of trajectory tracking or path following. The trajectory tracking problem involves the tracking of a timed reference position. A path-following approach eliminates any time dependency of the problem, which has many advantages for controlling performance and design \citep{sallouha2018energy}.
Considering path following control with minimum energy consumption, in \citep{gandolfo2016stable}, the authors examined the relationship between navigation speed and energy consumption in a miniature quadrotor helicopter, which travels over the desired path in an experimental study. Then a path-following controller proposed with a dynamic speed profile that varies with the geometric requirements of the path. The stability of the control law is proved using the Lyapunov theory.

\subsection{Battery-Aware \& Battery Performance}

It is not realistic to assume that the power drawn by a motor is in a 1:1 correspondence to the power drawn by the battery, since the battery supplies power with different efficiency values depending on its state-of-charge (SOC), and this efficiency is also non-linearly dependent on the amount of the power requested \citep{di2015energy, chen2006accurate}. Thus, omitting the battery performance analysis may result in inaccurate estimates of the real flight time of the drone \citep{aleksandrov2012energy}, and integrating battery awareness into the drone power model is essential to avoid significant errors \citep{chen2018case}. 

On the basis of empirical studies of battery usage for various UAV activities, \cite{abeywickrama2018empirical} presented a consistent model of power consumption for UAVs. In \citep{ahmed2016energy}, the experimental results were presented for a few basic UAV manoeuvring actions: hovering, flying vertically upward and flying vertically downward. In \citep{tseng2017flight}, the authors have investigated the impact of movement (hovering, vertical and horizontal movements), payload, and wind on the power consumption of an unmanned aerial vehicle.
\cite{abeywickrama2018comprehensive} proposed an enhanced energy consumption model and conducted a series of studies aimed at understanding the impact of several factors on UAV power consumption. A number of factors have been taken into consideration, including impact of wind, speed, tacking-off, hovering, payload, communication, and on-ground power consumption.
\cite{chen2018case} described a battery-aware model for assessing drone energy consumption, which is then applied to a scenario of drone delivery. According to the results, failing to account for battery performance leads to considerable inaccuracy in estimating the amount of available energy and, consequently, the duration of flight.

\cite{poikonen2017vehicle} presented a model for solving a problem of minimizing the delivery time for a certain number of packages. Battery performance was considered in this case solely from the viewpoint of service time.
\cite{chen2018case} described a battery-aware model for an accurate analysis of the drone energy consumption. This model was then applied to a scenario of drone delivery, and the results showed an accuracy more than about 16\% with respect to the traditional estimation model. \cite{baek2018battery} demonstrated that the proposed battery-aware delivery scheduling algorithm can carry more packages than the traditional delivery model with the same battery capacity. For the same delivery scheme, the battery-aware delivery model was 17\% more accurate than the traditional delivery model, which eliminates the possibility of a drone landing unexpectedly. As a result, the authors demonstrated how a model that incorporates SOC-dependent battery efficiency can be useful for modeling drone power. Using a battery-powered state-of-charge (SOC), a controller was described in \citep{podhradsky2014battery}, which can be used for both fixed wings and multirotor UAVs. In this scenario, the battery model was based on the equivalent electrical circuit of \citep{chen2006accurate} applied to LiPo batteries, as well as the relationship between nominal thrust and battery SOC.

\subsection{Target Tracking}

There are many situations in which visual tracking is used, such as search and rescue missions and the monitoring of vehicular traffic by tracking cars \citep{apvrille2014autonomous, heintz2007images}. A challenge of such a mission is the transmission of target images in real time, tracking the target accurately, and preserving the UAV's energy \citep{elloumi2017designing}. There are two phases in the tracking process. The transient phase of this process begins with the UAV taking off and localizing the mobile target. The second phase is known as the steady phase, which the UAV performs adjustments in order to maintain the target in its field of view. A fixed-wing UAV is tracking a target by making circular movements in \citep{he2014trackability}. When tracking a stationary target or when the target velocity is lower than the UAV's minimum velocity, the objective is to generate an optimal path for the UAV.

\cite{zorbas2013energy} illustrated a tracking of several targets by several unmanned aerial vehicles. The objective was to save energy while ensuring continuous coverage while only the altitude was adjusted and kept as low as possible during tracking. \cite{elloumi2017designing} proposed three zones single UAV tracking algorithm, and according to the target placement in those zones, the UAV will do a specific type of actions. In additional zone called the authorized zone, the UAV keeps a fixed velocity and a fixed altitude, and this contributes to the limitation of the energy consumption. The energy consumption was also evaluated using an adapted criterion, which takes into account the velocity, altitude, and acceleration changes. Limiting the UAV adjustments will reduce energy consumption while maintaining the target in sight.
\cite{siam2012board} introduced a method for multi-target detection and tracking based on fast corner detection and Kalman filtering, but the clustering algorithm was not able to identify the target. \cite{teuliere2011chasing} combined a particle filter algorithm with a color tracker based on multi-part representations in order to account for target occlusion and deformation.

\subsection{UAV-Enabled Mobile Edge Computing}

The UAV-assisted mobile edge computing (MEC) networks provide on-demand computation services to mobile terminals (MTs) through their high mobility and ease of deployment. There is a reduction in latency in this network, but energy efficiency remains a major concern, as both the UAVs and MTs have limited battery storage capacity \citep{budhiraja2021energy}.
It is difficult for user devices (UDs) to execute these applications on their own computing resources due to the limited battery power and low computational capacity \citep{hu2014energy}. As a solution, MEC offers UDs at the edge of wireless networks access to cloud computing services with a low transmission delay \citep{othman2013survey}. \cite{zhang2018stochastic} studied an energy minimization scheme for MEC networks based on UAVs to maximize computation rates. To accomplish this, the authors divided the primary problem into three subproblems: user scheduling, offloading ratio, and trajectory of the UAV.

\cite{sardellitti2015joint} examined a multicell MEC system where computation and radio resources were optimized together in order to minimize the total energy consumption. \cite{you2016energy} proposed an offloading policy aimed at reducing the energy consumption under the constraints of a data processing delay. \cite{zhou2018uav} investigated an MEC system with UAV assistance, in which the sum energy consumption at the UAV was minimized by optimizing the CPU frequency and trajectory of the UAV.
\cite{hua2018optimal} investigated the method for minimizing the energy consumption of a computation with a fixed trajectory for UAVs.
\cite{ji2020energy} aimed to formulate new optimization problems for non-orthogonal and orthogonal multiple access modes that minimize the weighted-sum energy consumption for the UAV and UDs by optimizing the UAV trajectory jointly with the allocation of computation resources, under the constraint of the number of computation bits.
\cite{budhiraja2021energy} evaluated how to minimize the energy consumption of NOMA-based MEC networks that support UAVs based on time, computation capacity, and trajectory. As a nonlinear programming problem, the formulated model is subdivided into two subproblems: joint time allocation and task computation capacity, and optimization of UAV trajectory.

\subsection{UAV-Enabled Multicasting}

It is possible to use UAVs as aerial base stations to improve the coverage and performance of communication networks in various scenarios, such as emergency communications and network access in remote locations \citep{liu2018energy}.

\cite{mozaffari2016unmanned} optimized the UAV stop points by using the disk covering theory. \cite{7881122} utilized a heuristic algorithm to optimize the 3D placement of multiple UAVs in order to serve all users. An optimal placement algorithm for UAV-BSs was proposed by \cite{shakhatreh2017efficient}, which maximizes the number of users covered with the least amount of transmission power possible. \cite{mozaffari2016efficient} developed a framework for determining the optimal location of UAVs based on circle packing theory in order to maximize the downlink coverage performance with minimal transmission power. \cite{song2019completion} proposed a fly-and-communicate protocol in which the UAV follows a zigzag trajectory rather than a spiral pattern trajectory optimization scheme and disseminates some common information to the ground units. In order to minimize the completion time, the optimal altitude is set as the maximum value, while the optimal beamwidth can be obtained through a one-dimensional search. The optimal altitude and beamwidth can be determined iteratively so as to minimize the energy consumption. \cite{yang2019three} aims to maximize the energy-efficient communication coverage of drone-cell networks while preserving the network connectivity with 3D continuous movement control of multiple drone cells. The E2CMC algorithm is based on an emerging deep reinforcement learning method to mitigate this problem. An energy efficiency reward function considering energy consumption, quality-of-service (QoS) requirements of users, and coverage fairness is first designed in E2CMC. As a result of interacting with an environment, E2CMC adjusts drone cells' locations continuously. If the networks are disconnected, E2CMC will reduce the reward drastically. \cite{liu2018energy} proposed leveraging emerging deep reinforcement learning (DRL) for UAV control, and present a novel, highly energy-efficient DRL-based method, entitled DRL-based energy efficient control for coverage and connectivity (DRL-EC3). The proposed method maximizes energy efficiency by taking into consideration communications coverage, fairness, energy consumption, and connectivity while learning the environment and its dynamics under the guidance of two deep neural networks. \cite{deng2019energy} investigated a novel UAV-enabled multicast system in which a UAV transmits CI to several GTs. To minimize the total energy consumption of the UAV, including mechanical and transmission energy, the objective is to minimize the total energy consumption of the UAV. An ML-based joint optimization framework for UAV-enabled multicasting is presented.

\subsection{Drone's Component Models}

An alternative approach for modeling drone energy consumption relies on a component model derived from helicopter operations, under the assumption that the power consumed during level flight, takeoff, or landing is approximately equivalent to the power consumed while hovering. Components can be modeled based on fundamental forces of flight, including the weight force of the aircraft (due to gravity) and drag force. Models for drone energy consumption include separate models for the forces and the different components of a drone flight (takeoff, landing, cruising, hovering), and are often quite detailed in order to capture particular characteristics of the drone design.


\cite{liu2017power} provided a detailed three-part drone energy model that includes the power to maintain lift and overcome parasite drag, along with profile power to overcome rotating drag caused by propeller blades. Field tests in \citep{liu2017power, di2015energy} showed that ascending takes 9.8\% more power than hovering, and descending takes 8.5\% less power than hovering. \cite{dorling2016vehicle} provided an equation for the power required by a multirotor helicopter in the hover mode as a function of battery capacity and payload weight. In \citep{kirschstein2020comparison}, a component model was used in an idealized delivery process with separate calculations for takeoff, ascent, level flight, descent, hovering, and landing.
\cite{stolaroff2018energy} used a required thrust to balance the drone weight and the parasitic drag force; the authors developed a two-component model; an assessment of the energy use and greenhouse gas (GHG) emissions for small drones with short ranges performing deliveries. Many articles have been developed the component drone power and energy models similar to those above for problems involving drones in wireless communication networks. (e.g., \cite{zeng2017energy, zeng2019energy, wu2019energy})
There is also a modeling of drone energy consumption that involves regression based on field experiments. \cite{tseng2017flight} and \cite{alyassi2017autonomous} presented a nine-term nonlinear regression model for the drone power consumption, which includes horizontal and vertical speed and acceleration, as well as payload mass and wind speed. A standard energy efficiency data set provided by manufacturers from an independent or governmental source would be ideal such as the energy efficiency measures for automobiles or appliances \citep{zhang2021energy}.

\subsection{Joined Models}

There are multiple uses for drones today, including emergency services for humanitarian operations (e.g., search and rescue) \citep{mudiyanselage2021automated,shakerian2022hybrid}, traffic surveillance, package delivery, and telemetry and mapping. A number of studies have been conducted that cover different aspects of the use of drones for modeling. \cite{yacef2017optimization} presented an energetic model for quadrotor UAVs, which contains the vehicle dynamic, actuator dynamic and battery dynamic with an efficiency function. The objective is to minimize the energy consumed by a quadrotor at the end of its mission while satisfying the boundary conditions and feasibility constraints on the states of the system and control inputs. \cite{yang2022global} proposed an energy consumption model for UAV swarm topology shaping that takes into account the energy consumption for UAVs flying vertically upward, vertically downward, and horizontally. \cite{van2021real} proposed a scheduling strategy by considering the UAV's characteristics in terms of energy consumption and reputation. For the scheduling strategy rules of UAVs, authors proposed an energy-efficient strategy and a reputation-based mechanism separately. Furthermore, they built a game-theoretic model to examine how working UAVs schedule new tasks. Finally, they calculated the Nash equilibrium for UAV scheduling based on the balance between energy minimization and reputation maximization. In \citep{tran2020coarse}, the UAV trajectory was designed to minimize the total energy consumption while meeting the requested timeout (RT) requirement and energy budget, which is achieved by optimizing the path and UAV's velocity along subsequent hops. Firstly, the authors proposed two algorithms, namely, heuristic search and dynamic programming (DP), to obtain a feasible set of paths without violating the GU's RT requirements based on the traveling salesman problem with time window (TSPTW). As a reference method, exhaustive search and traveling salesman problem (TSP) were compared. The results showed that the DP-based algorithm approaches the exhaustive search's performance with significantly reduced complexity.

However, the majority of studies focus on drone delivery. In the context of delivery services, i) meeting deadlines in terms of quality of service; ii) the number of packages delivered as measured by throughput per charge cycle; iii) improving battery health by reducing the number of charge cycles per time interval should be considered \citep{chen2018case}. \cite{bongermino2017model} presented a complete Simulink model and a control strategy for a parallel hybrid electric UAV powertrain. Model components include an internal combustion engine, a gearbox, an electric motor, an electric drive, and a lithium-ion battery. This control strategy employed a near real-time, iterative algorithm based on dynamic programming to solve an optimization problem involving optimal power management and torque-split for the powertrain with final state constraints.

The majority of studies on drone-only delivery systems assumed that there are several drones and each drone can serve one or more consumers every trip.
\cite{dorling2016vehicle} proposed two variations of the vehicle routing problem (VRP) for drone delivery. The first reduces overall operational costs while adhering to a delivery time limitation, whereas the second optimizes delivery time while adhering to a budget constraint. The prices include the costs of operating the drone fleet and the consumption of energy. To reduce costs, each drone may make several journeys and visit multiple consumers on each trip.
The authors used a linear approximation function that fluctuates linearly with the weight of the payload and battery rather than dealing directly with the original nonlinear form of the power function and solve the models with the simulated annealing (SA) heuristic. 
\cite{troudi2018sizing} analyzed an example of a drone delivery challenge with time constraints and a trip duration limit. Efforts are being made to reduce the number of drones used, the distance traveled, and the number of batteries required. Batteries are set aside as buffers for exceptional circumstances when applied linear energy limits.

Delivery can also be accomplished with a truck and one or more drones. There are several optimization models for drone or truck-drone routes or drone delivery systems that only indirectly consider energy consumption as a set constraint on drone endurance (flight time) or range (flight distance). (e.g., \cite{murray2020multiple, chiang2019impact, kitjacharoenchai2020two}. Others have used an energy consumption model based on the underlying physical forces involved in flight or field measurements in their drone delivery research (e.g., \cite{kirschstein2020comparison, murray2020multiple, poikonen2020multi, stolaroff2018energy, figliozzi2017lifecycle, dorling2016vehicle}). \cite{murray2020multiple} design truck-drone tandem delivery routes using a three-phase heuristic that considers multiple drone energy models, such as the model for fuel efficiency of \citep{liu2017power}, the simple regression model that is linear in payload, and other models that operate with a fixed distance or time limit (basically, modeling energy consumption as a linear function of drone travel time or distance). The findings from \citep{zhang2021energy} indicated that (i) different energy models can produce very different routes, with several energy models resulting in energy infeasible drone routes, and (ii) it is critical to include the energy consumed during steady level flight portions of a delivery trip (for example, for launch, retrieval, and delivery), especially for any hovering required to communicate with a truck or other drones prior to landing \citep{zhang2021energy}.

Drone energy consumption models can consist of only a few parameters or multiple interdependent components the provide accurate representations of flight forces and drone design. Since the seminal work of \citep{murray2015flying}, researchers have studied the possibilities of designing and optimizing drone delivery, in which drones are launched from a depot or other vehicle, which is usually a truck. This research includes drone routing and scheduling (e.g., \cite{dorling2016vehicle, agatz2018optimization, schermer2019hybrid, liu2019optimization, murray2020multiple, kitjacharoenchai2020two}), facility location problems, including charging stations \citep{chauhan2019maximum, hong2018range, ferrandez2016optimization}, and fleet sizing \citep{troudi2018sizing}. Recently conducted surveys of drone modeling specifically focus on truck-drone operations, in which drones can be transported by trucks to extend their useful range(or in delivery settings where trucks function as resupply points for drone deliveries (e.g., \cite{otto2018optimization, chung2020optimization}). Several other methods have been developed that enable drones to use public transportation in order to extend their useful delivery range \citep{huang2020reliable, huang2020new, choudhury2021efficient}.
In general, drone travel models impose time and/or distance limits as a result of their limited battery capacity.

The majority of the research assumes that the drone energy consumption is linear as a function of time or traveled distance, so drones are modeled as linear functions. However, there is considerable variance in the assumed consumption values. Drone energy consumption models have been incorporated explicitly into some optimization models, with one key difference being the assumption regarding thrust when flying horizontally. It is possible to assume (i) that the thrust force equals the drag force and that the weight force equals the lift force. Various assumptions are reflected by different perspectives on drone operations, for instance, whether they operate more like fixed-wing aircraft or helicopters. Based on these three approaches, there has been a continuous stream of literature on drone energy modeling.
\cite{d2014guest} aimed to a model drone energy consumption by translating the fundamental flight principles of manned aircraft into a model for unmanned aerial drones rather than manned aircraft. Using an integrated approach, this article presented a model that incorporates aerodynamics and drone design aspects into a single key parameter: the lift-to-drag ratio. Additionally, the energy model includes a fixed component for avionics power. \cite{troudi2018sizing} employ the same model to analyze drone fleet sizes; however they ignore the power of the avionics.
\cite{figliozzi2017lifecycle} adopted a same modeling approach to derive drone emissions based on a continuous approximation travel distance model. \cite{d2014guest} integrated a model that is also used in a series of reports from the RAND Corporation. The authors explored energy consumption for city-scale drone delivery systems \citep{lohn2017s, xu2017design, gulden2017energy}. \cite{lohn2017s} used this model to analyze truck and drone delivery in cities of various sizes, and \cite{gulden2017energy} provided a GIS-based analysis of shifting truck deliveries to drones in Minneapolis. \cite{xu2017design} discussed aspects of drone design related to drone energy consumption and suggested that fixed-wing VTOL (vertical takeoff and landing) or hybrid multicopter configurations that combine VTOL capabilities with lifting surfaces (wing-like structures) would be more suitable for many drone delivery purposes.

\bibliography{References-UAV.bib}

\end{document}